\documentclass[10pt, draftcls, onecolumn]{IEEEtran}
\IEEEoverridecommandlockouts
\usepackage{cite}
\usepackage{amsmath,amssymb,amsfonts}
\usepackage{algorithmic}
\usepackage{graphicx}
\usepackage{subfigure}
\usepackage{textcomp}
\usepackage{xcolor}
\def\BibTeX{{\mathrm B\kern-.05em{\sc i\kern-.025em b}\kern-.08em
    T\kern-.1667em\lower.7ex\hbox{E}\kern-.125emX}}
\usepackage{graphicx} 
\usepackage{epstopdf}

\usepackage{epstopdf}
\usepackage{array}
\usepackage{float}
\usepackage{subfigure}
\usepackage{color,cite,times}
\usepackage{CJK}
\usepackage{indentfirst}
\usepackage{amsmath}
\usepackage[T1]{fontenc}

\begin{document}
\title{Path-specific Underwater Acoustic Channel Tracking and its Application in Passive Time Reversal Mirror}

\author{
\IEEEauthorblockN{Xiuqing Li$^{\dagger}$, Wei Li$^{\dagger, \ddagger}$~\IEEEmembership{Member,~IEEE}, Xinlin Yi$^{\dagger}$, Qihang Huang$^{\dagger}$, Yuhang Wang$^{\dagger}$, Chenzhe Ye$^{\dagger}$}\\
\IEEEauthorblockA{$^{\dagger}$\textit{Harbin Institute of Technology (Shenzhen), China.} \\
$^{\ddagger}$\textit{Peng Cheng Laboratory, Shenzhen, China.} \\
\thanks{
This work is supported by National Natural Science foundation of China (61871151) , National Natural Science foundation of Guangdong Province (2018A030313177), and Guangdong Science and Technology Planning Project (2018B030322004), and the project ``The Verification Platform of Multi-tier Coverage Communication Network for oceans (LZC0020)''.}
\thanks{Wei Li is the corresponding author, email: li.wei@hit.edu.cn.}
}
}

\maketitle

\begin{abstract}
We consider the underwater acoustic channel which is  time-variant and doubly-spread in this work. 
Since conventional channel estimation and decision feedback equalizer (DFE) can not work well for this type of channel, a path-specific underwater acoustic channel tracking is proposed. 
It is based on the framework of Kalman filter. We provide a simplified sound propagation model as the state transition model.  A multipath tracker is proposed which is tolerant of the model-mismatch. Then we can obtain the time-variant path number and path-specific parameters such as delay and Doppler scaling factor. We also consider the application of the proposed  path-specific underwater acoustic channel tracking. 
We propose two types of passive time reversal mirror (PTRM) with our path-specific parameters for time-variant and doubly-spread underwater acoustic channel. With the path-specific parameters obtained by the proposed channel tracking, the proposed PTRM can not only match the time dispersion  as conventional PTRM, but also the doubly-spread channel, since the path-specific delay and Doppler scaler factor can  help to match the channel  in both time and frequency domain. For extensive doubly-spread channel, we can further apply the path-specific compensation to the PTRM.
 Both simulations and experimental results  by data from 2016  Qiandao Lake experiment   show the  efficiency of proposed path-specific channel tracking and proposed PTRMs with path-specific parameters.
\end{abstract}


\section{Introduction}


Underwater acoustic communications and signal processing depend on the ability of acquiring the underwater acoustic channel information. 
Channel estimation is a very popular topic for underwater acoustic communications in the last decade to estimate  the channel state information \cite{BZPW10}. 
However, in some rough sea, due to the currents, depth-changing, relative movement or  some other physical property of the oceans, the underwater acoustic channel can be severe doubly-spread, and fast time-variant. Since traditional channel estimation works  for  static state of channel, it is difficult to acquire time-variant  information of channel, hence limited.

Therefore, DFE often follows the channel estimation for time-variant channel.  DFE used in underwater channel  is usually  model-independent method such as the least mean square (LMS) \cite{Widrow} and the recursive least  squares (RLS) based method.  However, for the extensive time-variant and doubly-spread   underwater acoustic channel,  its performance is limited.

Model-based channel tracking  is introduced  for underwater acoustic channel recently,
and shows that the acquisition of underwater acoustic channel information is developed from channel estimation  to channel tracking.
It is usually based on the framework of Kalman filter.  A model is needed to provide  the time state transition for Kalman filter. 
\cite{Wei19} \cite{Huang14} \cite{TCYangicc}  take autoregressive (AR) Model as the  state model to the channel principle components based on the cross-correlation  property of underwater acoustic channel. 
With a Kalman filter applied on the AR model, it shows a better tracking ability than model-independent LMS and RLS. 
However, the AR model in \cite{Wei19} \cite{Huang14} \cite{TCYangicc}  is a static model which can not provide the transition  information of the time-variant channel.
Therefore, it limits the performance of the Kalman filter.

With the development of high-performance processors, more complex algorithm is accepted. The physical sound propagation model can be used in  channel tracking to  improve the performance. The physical propagation model can provide the state transition  information for the Kalman filter \cite{Wei18}. And the multipath can be taken as multi-target. Then multi-target tracking (MTT) methods can be used for multipath channel tracking. Broadly speaking there are three different approaches to MTT: multiple hypothesis tracking (MHT), joint probabilistic data association (JPDA), and random finite sets (RFS). 
Though the multipath physical propagation  model has some mismatch for underwater acoustic channel,
with the development of tracking theory, the advanced tracking  methods can be tolerant to the model-mismatch under Bayesian theory. \cite{Georgescu10} shows that the multipath information  can be potentially tracked by probability hypothesis density (PHD) \cite{PHD} filter and cardinalized probability hypothesis density (CPHD) \cite{CPHD} filter based on RFS theory.
Conventional channel estimation or tracking  can only obtain a general channel impulse response (CIR),
 but with the  MTT  methods, we can obtain path-specific  information of the multipath channel.

In this work, we first provide a path-specific channel tracking method.  A simplified physical propagation model is provided to describe the state transition of multipath varation. Then the  multi-Bernoulli (MB) filter  \cite{lu16} \cite{Karl16} \cite{Reuter} is used as the tracker for paths at each state.  
The multi-object particle multi-Bernoulli filter (MOP-MB) is originally an MTT method  belonging to RFS. The MTT problems work on estimating the number of targets and the kinematic state of each target. 
Therefore, the proposed  channel tracking can not only estimate the parameters of each path  at each state, but also the path number at each state.
Even though there may be some mismatches to the simplified physical propagation model, MB filter  can still perform as an efficient tracker. Besides, MB filer can track both the number and the state of the target with the initially unresolved measurements \cite{lu16}. Then even if the multipath becomes a clutter sometimes, MB filter can still track well. 


There are much potential use of the tracked time-variant path-specific parameters of underwater acoustic channel. They may improve the communication performance with more path-specific information of the channel other than  CIR in conventional channel estimation and tracking. It is not restricted to any special communication module.
This channel tracking can work on any kinds of modulation  if we apply it  at  the same stage as channel estimation. Besides, with the development of underwater systems, multi-purpose channel tracking is needed to help us explore the underwater environment, e.g. during communications, detection or localization tasks  \cite{Antonio} for autonomous underwater vehicle (AUV)s can be realized.

In this work, we consider to apply this  path-specific channel tracking to the communications with PTRM.
PTRM takes advantage of the heterogeneity of the medium to improve the focusing quality which could match the multipath acoustic channel and lead to a focusing. The property of temporal compression of PTRM reduces the dispersion  which means the inter-symbol interference is mitigated especially for large time dispersion, multipath underwater acoustic channel.


However, traditional PTRM \cite{TCYPTRM} \cite{TCSPTRM} \cite{Song} only considers the static time dispersion in mulipath channel.  It doesn't consider the spread in frequency or Doppler.  For doubly-spread channel, each path has different Doppler shift due to different physical propagation  \cite{Shengli2014}. 
Then conventional  PTRM can not match to this kind of channel, therefore lose efficacy on focusing.
With the help of  proposed MB path-specific channel tracking,  we propose two kinds of PTRMs for doubly-spread channel. The proposed PTRMs can match the multipath both in time and frequency domain with path-specific  parameters, therefore, improve the focusing.



The main contributions of this work:
\begin{itemize}
\item We propose a path-specific channel tracking for time-variant doubly-spread acoustic channel.
We provide state transition model based on a simplified sound propagation model. 
Parameters such as the depth of the transducer and the distance between the sender and receiver, usually considered in sound propagation model  with large observation error, are  avoided in this model. Then, under the Kalman filter framework, MB filter is applied to track the path-specific parameters and the path number. The proposed path-specific channel tracking  can  tolerate the mismatch of model.
 Therefore, it can be used for many underwater acoustic applications which need time-variant path-specific parameters.

\item  Then we propose two PTRMs with path-specific  channel tracking for time-variant doubly-spread acoustic channel.  
Our proposed PTRMs consider  the channel state in both time and frequency domain therefore can match to the doubly-spread channel. This can be done after we obtain the path-specific parameters of the channel with the proposed channel tracking. For the severe doubly-spread channel, we  propose a PTRM with a path-specific compensation  to improve focusing. 

\item  Both simulations and experiment results show the  efficiency of proposed path-specific channel tracking and PTRMs for time-variant doubly-spread underwater acoustic channel.
\end{itemize}

%
%

\section{Channel Model and Conventional PTRM}
\subsection{Doubly-spread Channel}

The received passband signal $y(t)$ is related to the transmitted passband signal $x(t)$ as
\begin{eqnarray}\label{y(t)}
 y(t) &=&x(t)* h(t;\tau)+n(t)\nonumber\\
&=&\int x(t-\tau)h(t;\tau){\mathrm d} \tau+n(t),
\end{eqnarray}
where $n(t)$ is noise, $*$ denotes the convolution operation.

We consider a time-variant underwater acoustic channel with ${N_{\mathrm{pa}}}$ multipath arrivals, each with an amplitude ${A_p}(t)$ and delay ${\tau _p}(t)$:
\begin{equation}\label{Dchannel}
h(t; \tau) = \sum\limits_{p = 1}^{{N_{\mathrm{pa}}}} {{A_p}(t)\delta (\tau  - {\tau _p}(t))}.
\end{equation}
We can assume that ${A_p}(t)$ and  ${\tau_p}(t)$ are slowly varying within a short time-block $T$, 
\begin{subequations}
\begin{align}  
{A_p}(t)  &= {A_p},   \\
{\tau _p}(t) &\approx {\tau _p} - {a_p}t,
\end{align} 
\end{subequations}
where ${\tau_p}$ is the initial delay and $-{a_p}$ is the first order derivative of ${\tau_p}(t)$. The parameter ${a_p}$ is often termed the Doppler scaling factor. Hence, we have a doubly-spread channel with path-specific Doppler scales as
\begin{equation}\label{Dchannel}
{h}(t;\tau) = \sum\limits_{p = 1}^{{N_{\mathrm{pa}}}} {{A_p}\delta (\tau  - {\tau_p} + {a_p}t)}.
\end{equation}



If we consider a time-variant channel with a much longer duration,  the channel is divided into  blocks with $T$-length, then within each block,  \eqref{Dchannel} still holds. The whole doubly-spread  time-variant channel  is described as follow
\begin{equation}\label{Dtchan}
\quad\quad  {h}(t;\tau) = \sum\limits_{p= 1}^{{N_{\mathrm{pa}}}_k } {{A^p_k }\delta (\tau - {\tau^p_k } + {a^p_k }t)}, \quad     \quad\quad  (k-1)T<t<kT,\quad k=1, 2, \cdots, \infty,
\end{equation}
where $A^p_k$, $\tau^p_k$ and $a^p_k$ are the amplitude, delay and Doppler scaling factor for the $p$th path  during $k$th time block, respectively, and ${N_{\mathrm{pa}}}_k$ is the multipath arrival number  during $k$th time block. It is assumed these parameters do not change within the time block, but may change from block to block for time-variant channel.
Then for the doubly-spread  time-variant channel as \eqref{Dtchan}, the received signal is related to the transmitted signal in passband  as
\begin{equation}\label{y(t)tv}
\quad\quad  y(t)=\sum\limits_{p = 1}^{{N_{\mathrm{pa}}}_k} {{A_k^p} x([1 + {a_k^p}]t - {\tau_k^p})}  +  w(t), \quad \quad \quad  (k-1)T<t<kT,\quad k=1, 2, \cdots, \infty,
\end{equation}
where $w(t)$ is additive noise with model-mismatch and ambient noise.

For time-invariant channel, all the paths are stable with no delay variations. 
The channel is usually simplified as
\begin{equation}\label{Schannel}
h(\tau) = \sum\limits_{p = 1}^{{N_{\mathrm{pa}}}} {{A_p}\delta (\tau  - \tau _p)}.
\end{equation}

Generally, conventional PTRM is based on \eqref{Schannel}.  
Obviously, this simplified version of channel model  may not work if the channel is severe  doubly-spread and time-variant.

\subsection{Conventional PTRM}

In conventional PTRM,
we assume if we have time-reversed CIR $h(-\tau)$  for time-invariant channel  \eqref{Schannel}, then
\begin{eqnarray}
 z(t) &=& {h(-\tau)*}{y}(t)\nonumber\\
&=& h(-\tau)*h(\tau)*x(t)\nonumber\\
 &= & q(\tau) *x(t),
\end{eqnarray}
where $*$ is  defined as  \eqref{y(t)}, and $q(\tau)$ is time reversal acoustic field, or $Q$-function(QF)
\begin{equation}
\label{qfunc}
q(\tau) =  h(-\tau)* h(\tau).
\end{equation}
Clearly, for time-invariant channel as in \eqref{Schannel}, QF $q(\tau) \approx  \sum_{p= 1}^{N_{\mathrm{pa}}} {A_p}^2\delta(\tau)$. Array processing is often applied to strengthen the delta function.

In practice of conventional PTRM, a probe signal such as a chirp signal is received in advance of the data packet and is used to correlate with the received data to estimate the channel-impulse response $\hat h(\tau)$,  and  $\hat h(-\tau)$ is obtained by simply flipping the estimate of the CIR $\hat h(\tau)$ in time domain.  If Doppler can not be avoided, we obtain an estimation of one general overall Doppler shift, 
and Doppler compensation with this Doppler shift  is applied before PTRM to decrease its effect. 
An exponentially weighted RLS algorithm is often applied to update the DFE tap weights. The residual carrier phase offset in $z(t)$ is compensated by a second-order phase locked loop in the adaptive DFE. This DFE may help  for slow time-variant channel.

\section{MB based Channel tracking}

In this work, we consider \eqref{y(t)} and \eqref{Dtchan}  as the channel model where the input-output relationship is parameterized by ${N_{\mathrm{pa}}}_k$ triplets $\{ {A_k^p},{\tau_k^p},{a_k^p}\}_{p=1}^{{N_{\mathrm{pa}}}_k} $ for each time state $k$.

To work on the time-variant channel, we propose a channel tracking method based on MOP-MB filter. 

An MOP-MB filter recursively computes and propagates the probability density function (pdf) of multi-object state in time, via Bayesian prediction and update steps.  It is based on Kalman filter framework. A  tracker based on Kalman filter needs: a state transition model and an observation model.  We first demonstrate  the details of these two models for multipath channel tracking and then the path-specific channel tracking based on the MB tracker.

\subsection{State Transition Model for multipath tracking}

Since ${\hat {A_k^i}}$  can be obtained due to the delay ${\hat \tau_k^i}$ from the measurements,
we develop a transition model  for delay and Doppler scaling factor.
The state vector for the $i$th path at time $k$ consists of delay and Doppler scaling factor 
\begin{equation}\label{statevector}
\mathbf{x}_k^i = \left[
\begin{array}{lcr}
\tau_{k}^i \\
a_k^i
\end{array}
\right],
\end{equation}
and let the multitarget state set be denoted
\begin{equation}
\label{eqn_multitargetstate}
\mathbf{X}_k=\{ \mathbf{x}_k^i \}_{i=1}^{N_k^{x}},
\end{equation}
where $N_k^{x}$ is the number of potential paths at time $k$.

\begin{figure}[htbp]
	\centering
	\includegraphics[angle=0,width=0.6\textwidth]{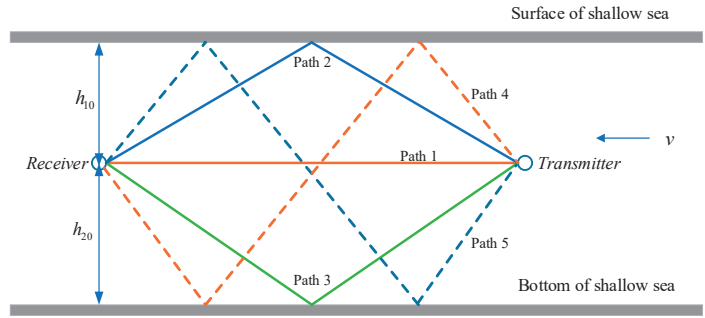}
	\caption{A simplified  underwater multipath model for shallow water environment. }
   \label{fig:channnel1}
\end{figure}

\begin{figure}[htbp]
	\centering
	\includegraphics[angle=0,width=0.45\textwidth]{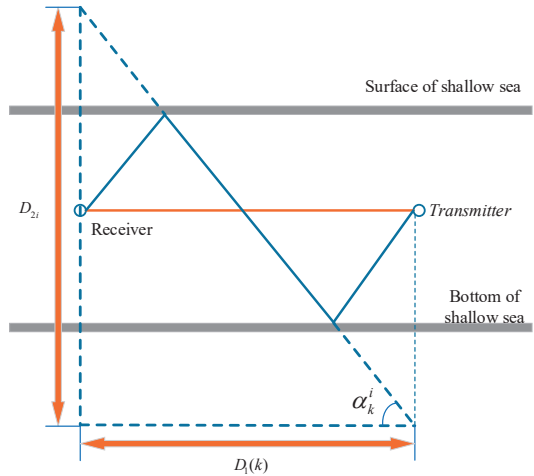}
	\caption{Path-5 is taken as an example to show the  equivalent path length in Fig.\ref{fig:channnel1}.}
   \label{fig:channnel2}
\end{figure}

In shallow water where there is no much temperature difference, the traces of paths can be taken as mirror reflection. 
A simplified ray-tracing underwater multipath channel model is shown in Fig.\ref{fig:channnel1}. 
It is a simplified version of ray-tracing sound propagation model by assuming the temperature does not change with depth too much, therefore the sound speed does not change with depth too much. It is reasonable for shallow water. And since the proposed tracking method is tolerant to the model-mismatch, this assumption is acceptable.
Assume some initial channel physics between source and receiver: the depth of receiver  $h_{10}$, the distance to the bottom  $h_{20}$, the relative speed of transmitter to  receiver $v$, initial distance between the transmitter and the receiver ${d_{\mathrm{SR}}}$, sound speed $c$ and observation interval $T$.  Fig.\ref{fig:channnel2} shows the equivalent multipath model with the equivalent horizontal distance ${D_1}(k)$ at state time $k$, the  equivalent vertical depth ${D_{2i}}$ for the $i$th path and angle of reflection $\alpha _k^i$.

The delay of the $i$th path at state time $k$ in \eqref{statevector} is given by
\begin{equation}
\label{tauk}\tau _k^i = \frac{{\sqrt {D_i^2(k) + D_{2i}^2} }}{c},
\end{equation}
and the Doppler scaling factor  in \eqref{statevector} is
\begin{equation}
\label{dopplerk}
a_k^i = \frac{v}{c} \times \cos \alpha _k^i = \frac{{v \times {D_i}(k)}}{{{c^2} \times \tau _k^i}}.
\end{equation}

From Eq.(\ref{tauk}) and Eq.(\ref{dopplerk}), we obtain
\begin{IEEEeqnarray}[\setlength{\nulldelimiterspace}
{2pt}]{rl's}\label{lengthk}
&{{D_i}(k) = \frac{{{c^{\mathrm{2}}}\tau _k^ia_k^i}}{v}},\IEEEyesnumber\IEEEyessubnumber\\*
[-0.125\normalbaselineskip]
\smash{\left\{\IEEEstrut[6\jot][6\jot]\right.}&&
\nonumber\\*[-0.125\normalbaselineskip]
&{{D_{{\mathrm{2}}i}} = \sqrt {{c^{\mathrm{2}}}{{(\tau _k^i)}^2} - \frac{{{c^4}{{(\tau _k^i)}^2}{{(a_k^i)}^2}}}{{{v^2}}}} }.\IEEEyessubnumber
\end{IEEEeqnarray}

Then the recurrence relation of the state from  $\mathbf{x}_{k}^i$ to  $\mathbf{x}_{k+1}^i$  is as follow
\begin{align}\label{unlifun}
 \mathbf{x}_{k+1}^i &= f_{k+1,k}(\mathbf{x}_k^i)  \nonumber\\
&= \left[\begin{IEEEeqnarraybox*}[][c]{,c,}
\frac{\sqrt{v^2T^2+c^2(\tau_k^i)^2+2c^2a^i_k(\tau_k^i)T}}{c}\\
\frac{c^2a^i_k\tau^i_k+v^2T}{c\sqrt{v^2T^2+c^2(\tau^i_k)^2+2c^2a^i_k\tau^i_kT}}
\end{IEEEeqnarraybox*}\right].
\end{align}
Consider the process noise, we have the state transition model  as follow
\begin{equation}
\label{eqn_statefunction}
 \mathbf{x}_{k+1}^i = f_{k+1,k}(\mathbf{x}_k^i) + \mathbf{v}_{k},
\end{equation}
where $\mathbf{v}_{k}$ is zero-mean white Gaussian process noise with covariance $\mathbf{Q}_k$.

Note that, the state equation is applicable  to paths with arbitrary times of reflection.

\subsection{Observation Model for multipath tracking}

For channel tracking, the set of multipath measurements is  denoted as $\mathbf{Z}_k$,
\begin{equation}
\label{measset}
\mathbf{Z}_k=\{\mathbf{z}_k^j\}_{j=1}^{N_k^z},
\end{equation}
where  $N_k^z$ represents the number of measurements at time $k$. $\mathbf{z}_k^j$ is a measurement from the $j$th path at time $k$,
\begin{equation}
\mathbf{z}_k^j = \left[
\begin{array}{lcr}
\tilde{\tau}_{k}^j \\
\tilde{a}_k^j
\end{array}
\right],
\end{equation}
where $\tilde{\tau}_{k}^j$ and $\tilde{a}_k^j$ are the estimations of  $\tau_{k}^j$ and $a_k^j$ which can be obtain according to \cite{Sharif} and \cite{MHFM}.
The observation model at time $k$ can be approximated by
\begin{equation}
\label{eqn measurementequation}
\mathbf{z}^j_{k}=\mathbf{H}\mathbf{x}^j_{k}+\mathbf{w}_{k},
\end{equation}
where $\mathbf{H}$ is an  identity matrix, and $\mathbf{w}_{k}$ is a white Gaussian random vector with measurement error covariance $\mathbf{R}_{k}$.

The  measurements are supposed to be the union of the sets of aforementioned measurements (\ref{eqn measurementequation}) generated from real paths ${ \mathbf{W}}_k(\mathbf{x}^i_k)$ and other clutter measurements ${\mathbf{C}}_k$:
\begin{equation}
\label{eqn_multitargetmeaset}
\mathbf{Z}_k={ \mathbf{C}}_k \cup \left[ \bigcup_{i=1}^{N_k^x} { \mathbf{W}}_k(\mathbf{x}^i_k) \right].
\end{equation}

The clutter measurement set is usually modeled as poisson process that each one is uniform distributed in the surveillance area with the pdf $g_{\mathrm c}(\mathbf{z})$ and the number of them is poisson distributed with parameter $\lambda_{\mathrm c}$,  $N_k^c$ represents the number of clutters at time $k$. The clutter set pdf is \cite{R. Mahler}
\begin{equation}\label{clutterpdf}
\kappa ({{\mathbf{C}}_k}) = {e^{ - {\lambda _c}}}\mathop \Pi \limits_{j = 1}^{N_k^c} {\lambda _c}{g_c}({\mathbf{z}}_k^j).
\end{equation}

\subsection{MB tracker for channel tracking}
After we obtain the observation model \eqref{eqn_multitargetmeaset} and the state transition model  \eqref{eqn_statefunction}, we can apply our MOB-MB  tracker which is based on Kalman filter to track the delay and Doppler scaling factor.  We assume each $\mathbf{x}_k^i$ for the $i$th path is an MB component   and is Gaussian distributed, then each  MB component density for $\mathbf{x}_k^i$ is described  by a mean $\mathbf{m}_k^i$, a covariance $\mathbf{P}_k^i$, and a weight $w_k^i$.  The MOP-MB filter has  four main steps: birth, prediction, update, and confirmation,  and through the four main steps we track the $\mathbf{m}_k^i$, $\mathbf{P}_k^i$, and $w_k^i$  and estimate the ${N_{\mathrm{pa}}}_k$ for each time state, therefore obtain the tracked $\mathbf{x}_k^i$ for every path.

Then the MB probability density of the multpath state set \eqref{eqn_multitargetstate} can be abbreviated as
\begin{equation}\label{mbpdf}
f({\mathbf{X}_k}) = {\{ (w_k^i,\mathbf{m}_k^i,\mathbf{P}_k^i)\} _i}.
\end{equation}
The explicit expression of \eqref{mbpdf} is in \cite{lu16} \cite{Karl16}.

\subsubsection {Initialization and path birth}

Every measurement obtained from the first step is regarded as potential eigenpath. Hence, all the measurements are directly loaded as the MB components at the first time. Similarly, for subsequent time steps, $N^{\mathrm{b}}_{k-1}$ measurements not associated with any current MB components can be potential new paths including both path birth and path spawning. The Gaussian MB density representing newborn paths at time $k-1$ is
\begin{equation}
\{ ({w^{\mathrm{b}}},\mathbf{m}_{k-1}^{\mathrm{b}(i)},{\mathbf{P}^{\mathrm{b}}})\} _{i = 1}^{N_{k-1}^{\mathrm{b}}},
\end{equation}
where   ${\mathbf{m}}_{k-1}^{\mathrm{b}(i)} = {\mathbf{z}}_{k-1}^{\mathrm{b}(i)} $  is set according to the measurements with no existing MB component, the existence probability ${w^{\mathrm{b}}}$ and the covariance ${\mathbf{P}^{\mathrm{b}}}$ are equal for all paths and can be suitably user-defined.

\subsubsection {Prediction}

We can obtain the updated MB components at time $k-1$ as
\begin{equation}
\label{eqn_example}
\{ (w_{k-1|k-1}^i,\mathbf{m}_{k-1|k-1}^i,\mathbf{P}_{k-1|k-1}^i)\} _{i = 1}^{N_{k-1|k-1}^{{\mathrm{MB}}}},
\end{equation}
in which $N^{\mathrm{MB}}_{k|k}$ is the number of components. Then the number of components is added by birth components
\begin{equation}
\label{eqn_example}
N^{\mathrm{MB}}_{k|k-1}=N^{\mathrm{MB}}_{k-1|k-1}+N^{\mathrm b}_{k-1}.
\end{equation}

The Gaussian MB component densities are predicted with the Kalman Filter one-step prediction formula as
\begin{subequations}
\begin{align}
w^i_{k|k-1}&=p_{\mathrm s}w^i_{k-1|k-1},  \\
\mathbf{m}_{_{k|k-1}}^i&={f_{k,k-1}}(\mathbf{m}_{k-1|k-1}^i), \\
\mathbf{P}_{_{k|k-1}}^i&={\mathbf{F}_{k-1}}\mathbf{P}_{k-1|k-1}\mathbf{F}_{k - 1}^{\prime} + {\mathbf{Q}_{k-1}}, 
\end{align}
\end{subequations}
where $p_{\mathrm s}$ is the probability of survival. ${\mathbf{F}_{k-1}}$ is the Jacobian matrix derived from $f_{k,k-1}(\cdot)$  as in \eqref{unlifun}. ${\mathbf{Q}_{k-1}}$ is the process noise covariance. Hence, the predicted MB density for each path is
\begin{equation}
\label{eqn_example}
\{ (w_{k|k-1}^i,\mathbf{m}_{k|k-1}^i,\mathbf{P}_{k|k-1}^i)\} _{i = 1}^{N_{k|k-1}^{{\mathrm{MB}}}}.
\end{equation}

\subsubsection {Update}

Update steps can be realized by MOP filter. It has three main steps: First, the pdf of predicted MB components is replaced by a particle approximation of the predicted multipath density $f({\mathbf{X}_k}|{\mathbf{Z}^{k - 1}})$. Next, each multi-object particle is updated by the multi-object update with an approximation for data association. It is a simplified processing which uses an auction algorithm rather than JPDA methods. Finally the posterior MB density is obtained approximately with all the posterior multi-object particles.

 (1) {A particle approximation of the predicted multipath density}
 
The pdf of predicted MB components is replaced by $M$ multi-object particles 
${\mathbf{X}_{k|k-1}^l}$ approximation. The predicted multipath density is approximated by 
\begin{equation}
\label{eqn_CK}
f({\mathbf{X}_k}|{\mathbf{Z}^{k - 1}}) \approx \sum\limits_{l = 1}^M {W_{k|k - 1}^l\phi_{\mathbf{x}_k}(\mathbf{X}_{k|k - 1}^l)}, 
\end{equation}
where ${W_{k|k - 1}^l}$ is the prior weight of each MOP, and ${\mathbf{Z}^{k - 1}}=\{\mathbf{Z}_\kappa\}_{\kappa=1}^{k-1}$ is the set of all measurements up to time $k-1$, and the MOP joint density is
\begin{equation}
\label{MOPdensity}
\begin{array}{l}
\phi_{\mathbf{x}_k}(\mathbf{X}_{k|k - 1}^l)\\
{\mathrm{ = }}\prod\limits_{i \in {I_l}} {N({\mathbf{x}_i};\mathbf{m}_{k|k - 1}^i,\mathbf{P}_{k|k - 1}^i)} \delta \left[ {\left| {{\mathbf{X}_k}} \right| - \left| {\mathbf{X}_{k|k - 1}^l} \right|} \right],
\end{array}
\end{equation}
where $|\mathbf{X}_k|$ represents the cardinality of  the set  $\mathbf{X}_k$, and the set ${\mathbf{I}_l}$ demonstrates the existence of each MB component for the $l$th multi-object particle particle. The $i$th entry of set ${\mathbf{I}_l}$ is defined as
\begin{equation}
I_l^i = \left\{ {\begin{array}{*{20}{c}}
1&{\begin{array}{*{20}{c}}
{if \ u_l^i \le w_{k|k - 1}^i}
\end{array}}\\
0&{otherwise},
\end{array}} \right.
\end{equation}
where  $ u_l^i$ is obtained by random sampling from the uniform distribution $U(0,1)$. Thus, the meaning of the set ${\mathbf{I}_l}$ is that the $i$th predicted Gaussion component is included with probability $w_{k|k - 1}^i$  in the $l$th multi-object particle.

(2) Each multi-object particle is updated through the multi-object update with an approximation for data association

The posterior multi-object distribution based on all observations up to time $k$ is given by a Bayesian update with a set integral, and with \eqref{eqn_CK}
\begin{equation}
\label{posterior}
\begin{array}{ll}
f(\mathbf{X}_{k}|\mathbf{Z}^{k})&=\cfrac{f(\mathbf{Z}_{k}|\mathbf{X}_{k})f(\mathbf{X}_{k}|
\mathbf{Z}^{k-1})}{\int f(\mathbf{Z}_{k}|\mathbf{X}_{k})f(\mathbf{X}_{k}|\mathbf{Z}^{k-1})\delta \mathbf{X}_{k}},\\
&=\cfrac{\sum_{l=1}^M f(\mathbf{Z}_{k}|\mathbf{X}_{k})
\phi_{\mathbf{X}_{k}}(\mathbf{X}_{k|k-1}^l)
}{\sum_{l=1}^M \int f(\mathbf{Z}_{k}|\mathbf{X}_{k})\phi_{\mathbf{X}_{k}}(\mathbf{X}_{k|k-1}^l)\delta \mathbf{X}_{k}},
\end{array}
\end{equation}
where $f(\mathbf{Z}_{k}|\mathbf{X}_{k})$ is the likelihood function of the multipath measurement set, and the integral is defined as in \cite{Karl16}. 
 According to \cite{R. Mahler}, we assume that present paths are detected with probability $p_{\mathrm D}$, and detections from present paths have Gaussian measurement error $ {g_x}(\mathbf{z}|\mathbf{x})$. Under these assumptions and combining the clutter set pdf in \eqref{clutterpdf}, the measurement set density is given as
\begin{equation}
\label{measurementpdf}
\begin{array}{l}
f({\mathbf{Z}_k}|{\mathbf{X}_k}) = \sum\limits_{\theta  \in \mathbf{\Theta} } {{e^{ - {\lambda _c}}}\left[ {\prod\limits_{j:\forall {\sigma _i} \ne j} {\frac{{{\lambda _c}}}{V}} } \right]} \\
 \times \left[ {\prod\limits_{i :{\sigma _i} = 0} {(1 - {p_{\mathrm D}})} } \right] \times \left[ {\prod\limits_{i :{\sigma _i} > 0} {{p_{\mathrm D}}{g_x}(\mathbf{z}_k^{{\sigma _i}}|\mathbf{x}_k^i)} } \right].
\end{array}
\end{equation}
Note that $\theta  = \{ {\sigma _i}\}$ represents data associations. Here ${\sigma _i} = 0$ if the path state ${\mathbf{x}}_k^i$ is not associated to any measurement, and ${\sigma _i} = j$ if ${\mathbf{x}}_k^i$ is associated to measurement ${\mathbf{z}}_k^j$.
Using the measurement set pdf \eqref{measurementpdf} and the MOP joint density \eqref{MOPdensity}, for each multi-object particle we have
\begin{equation}\label{MOPpdf}
\begin{array}{l}
f({\mathbf{Z}_k}|{\mathbf{X}_k})\phi {\mathbf{x}_k}(\mathbf{X}_{k|k - 1}^l)\\
 = \sum\limits_{\theta  \in \mathbf{\Theta} } {{e^{ - {\lambda _c}}}\left[ {\prod\limits_{j:\forall {\sigma _i} \ne j} {\frac{{{\lambda _c}}}{V}} } \right]}  \times \left[ {\prod\limits_{i:{\sigma _i} = 0} {(1 - {p_{\mathrm D}})} } \right]\\
 \times \left[ {\prod\limits_{i:{\sigma _i} > 0} {{p_{\mathrm D}}N(\mathbf{z}_k^{{\sigma _i}};\mathbf{\hat z}_k^i,\mathbf{S}_k^i)} } \right]\\
 \times \left[ {\prod\limits_{i \in {\mathbf{I}_l}} {N(\mathbf{x}_{k|k - 1}^i;\mathbf{m}_{k|k - 1}^i,\mathbf{P}_{k|k - 1}^i)} } \right]\\
 = \sum\limits_{\theta  \in \mathbf{\Theta} } {L_{_{k|k - 1}}^{l,\theta }\prod\limits_{i \in {\mathbf{I}_l}} {N(\mathbf{x}_k^i;\mathbf{m}_{k|k - 1}^{i,{\sigma _i}},\mathbf{P}_{k|k - 1}^{i,{\sigma _i}})} } \\
 = \sum\limits_{\theta  \in \mathbf{\Theta} } {L_{_{k|k - 1}}^{l,\theta }\phi {\mathbf{x}_k}(\mathbf{X}_{k|k - 1}^{l,\theta })} .
\end{array}
\end{equation}
The likelihood of the association event $\theta$ in the $l$th MOP is given by
\begin{equation}
\begin{array}{l}
L_{_{k|k - 1}}^{l,\theta } = {e^{ - {\lambda _c}}}{\left( {\frac{{{\lambda _c}}}{V}} \right)^{{N_{FA}}(\theta )}}\left[ {\prod\limits_{i:{\sigma _i} = 0} {(1 - {p_{\mathrm D}})} } \right]\\
 \times \left[ {\prod\limits_{i:{\sigma _i} > 0} {{p_{\mathrm D}}N(\mathbf{z}_k^{{\sigma _i}};\mathbf{\hat z}_k^i,\mathbf{S}_k^i)} } \right],
\end{array}
\end{equation}
where ${{N_{FA}}(\theta )}$ is the number of measurements that are not associated to a path state. Then we can update the MB component - the $i$th path by the extended Kalman filter (EKF) at time $k$:
\begin{equation}
\label{eqn_usp}
{\mathbf{m}}_{k|k}^{i,{\sigma _i}}={\mathbf{m}}^i_{k|k-1}+\mathbf{K}_{k}^i(\mathbf{z}_{k}^{\sigma _i}-\widehat{\mathbf{z}}
_{k}^i),
\end{equation}
and the covariance:
\begin{equation}
\label{eqn_prd2}
\mathbf{P}_{k|k}^{i,{\sigma _i}}=\mathbf{P}_{k|k-1}^i-\mathbf{K}^i_{k}\mathbf{S}_{k}^i(\mathbf{K}^{i}_{k})^{\prime},
\end{equation}
where ${\mathbf{m}}_{k|k}^{i,{\sigma _i}}={\mathbf{m}}_{k|k}^i$ and $\mathbf{P}_{k|k}^{i,{\sigma _i}}=\mathbf{P}_{k|k}^i$ for the special case of  ${\sigma _i} = 0$. 
The estimated measurement $\widehat{\mathbf{z}}_{k}^i$ in (\ref{eqn_usp}) is derived from ${\mathbf{m}}_{k|k-1}^i$ according to (\ref{eqn measurementequation}) as
$\widehat{\mathbf{z}}_{k}^i=\mathbf{H}{\mathbf{m}}_{k|k-1}^i. $
The gain of the EKF , $\mathbf{K}_{k}^i$, in (\ref{eqn_usp}) can be calculated by
\begin{equation}
\label{eqn_example}
\mathbf{K}_{k}^i=\mathbf{P}_{k|k-1}^i\mathbf{H}(\mathbf{S}_{k}^i)^{-1},
\end{equation}
where
\begin{equation}
\label{eqn_example}
\mathbf{S}_k^i=\mathbf{H}\mathbf{P}_{k|k-1}^i\mathbf{H}^{\prime}+\mathbf{R}_{k},
\end{equation}
in which $\mathbf{R}_{k}$ is the covariance of measurement noise.

Note that \eqref{MOPpdf} includes a summation over $\mathbf{\Theta}$, the set of all possible measurements associations $\theta$. Because of the computational complexity of the data association problem, it is generally considered infeasible to implement. 
To mitigate computational complexity, the auction algorithm \cite{D. P. Bertsekas} is used to compute the single most probable association event ${\hat \theta (l)}$ for each MOP. Under this approximation, the Bayesian normalization constant $f({\mathbf{Z}_k}|{\mathbf{Z}^{k - 1}})$ - the set integral in the denominator of \eqref{posterior} becomes
\begin{equation}
\int {f({\mathbf{Z}_k}|{\mathbf{X}_k})\phi {\mathbf{x}_k}(\mathbf{X}_{k|k - 1}^l)} \delta {\mathbf{X}_k} = 
L_{_{k|k - 1}}^{l,\hat \theta (l)}\delta \left[ {\left| {\mathbf{X}_k^{}} \right| - \left| {\mathbf{X}_{k|k - 1}^l} \right|} \right].
\end{equation}

We, thus, use the posterior multi-object particles to approximate the posterior multi-object density \eqref{posterior} as
\begin{equation}
\begin{array}{ll}
f({\mathbf{X}_k}|{\mathbf{Z}^k})&= \cfrac{{\sum\nolimits_{l = 1}^M {L_{k|k - 1}^{l,\hat \theta (l)}} \phi {\mathbf{x}_k}(\mathbf{X}_{k|k}^{l,\hat \theta (l)})}}{{\sum\nolimits_{l{\mathrm{ = }}1}^M {L_{k|k - 1}^{l,\hat \theta (l)}} }}\\
 &= \sum\limits_{l = 1}^M {W_{k|k}^l\phi {\mathbf{x}_k}(\mathbf{X}_{k|k}^{l,\hat \theta (l)})} ,
\end{array}
\end{equation}
where 
\begin{equation}
W_{k|k}^l = \frac{{L_{k|k - 1}^{l,\hat \theta (l)}}}{{\sum\nolimits_{l{\mathrm{ = }}1}^M {L_{k|k - 1}^{l,\hat \theta (l)}} }}.
\end{equation}

(3) Approximate Posterior multi-Bernoulli density

Since each MOP in a gate group has its own data association, a predicted estimate may be included in multiple MOPs,  and it follows that there may be multiple updated estimates that correspond to the same predicted estimate. We merge the results given by each MOP to obtain
\begin{equation}
f(\mathbf{X}_{k|k})=\{ (w_{k|k}^i,\mathbf{m}_{k|k}^i,\mathbf{P}_{k|k}^i)\} _{i = 1}^{N_{k|k}^{{\mathrm{MB}}}},
\end{equation}
where
\begin{subequations}
\begin{align}
w_{k|k}^i &= \sum\limits_{l:i \in {\mathbf{I}_l}} {W_{k|k}^l}, \\
\mathbf{m}_{k|k}^i &= \frac{1}{{w_{k|k}^i}}\sum\limits_{l:i \in {\mathbf{I}_l}} {W_{k|k}^l\mathbf{m}_{k|k}^{i,{{\hat \sigma }_i}(l)}}, \\
\mathbf{M}_{k|k}^{i,{{\hat \sigma }_i}(l)} &= (\mathbf{m}_{k|k}^{i,{{\hat \sigma }_i}(l)} - \mathbf{m}_{k|k}^i){(\mathbf{m}_{k|k}^{i,{{\hat \sigma }_i}(l)} - \mathbf{m}_{k|k}^i)^{\prime}},\\
\mathbf{P}_{k|k}^i &= \frac{1}{{w_{k|k}^i}}\sum\limits_{l:i \in {\mathbf{I}_l}} {W_{k|k}^l(\mathbf{P}_{k|k}^{i,{{\hat \sigma }_i}(l)} + \mathbf{M}_{k|k}^{i,{{\hat \sigma }_i}(l)})}.
\end{align}
\end{subequations}
\subsubsection {Pruning, confirmation and extraction}
At the last step of each  iteration, threshold  to the probability of existence $w^i_{k|k}$ is used to make a decision and extract the estimated results of multipath: MB components. If $w^i_{k|k}$ is  lower than a threshold, $\tau_{\mathrm p}$ will be pruned; ones with $w^i_{k|k}$ larger than a threshold $\tau_{\mathrm c}$ are confirmed as estimated eigenpaths; ones with $w^i_{k|k}$  larger than a threshold $\tau_{\mathrm e}$ are taken as the existence of estimated eigenpaths at time $k$. 

Then the estimated number of these eigenpaths at time $k$ is $\hat N_{\mathrm{pa}_k}$.
And the tracking results $\hat{\mathbf{x}}_k^i=\mathbf{m}_k^i$, therefore we have ${\hat \tau _k^p}$ and ${\hat a_k^p}$. 
 With the estimated ${\hat A_k^p}$ from measurements,
 we obtain $\{ {\hat A_k^p},{\hat \tau _k^p},{\hat a_k^p}\} _{p = 1}^{\hat N_{\mathrm{pa}_k}}$ after the MB based channel tracking.


\section{Path-specific PTRM for Doubly-spread time-variant Channel }

After we obtain  $\{ {\hat A_k^p},{\hat \tau _k^p},{\hat a_k^p}\} _{p = 1}^{\hat N_{\mathrm{pa}_k}}$ for each time $k$ through the above path-specific channel tracking based on MB tracker.  The tracked channel is as follow, 
\begin{equation}\label{est_channel}
{\hat h}(t; \tau) = \sum\limits_{p = 1}^{{{N}_{pa}}_k} {{\hat A_k^p}\delta (\tau  - {\hat \tau _k^p} + {\hat a_k^p}t)}, \quad\quad  (k-1)T<t<kT,\quad k=1, 2, \cdots, \infty.
\end{equation}

In this section, we propose two types of path-specific PTRMs in the following for fast time-variant  multipath channel.
We first propose a path-specific  PTRM with our path-specific channel tracking results.
Then we propose a  PTRM with path-specific compensation on Doppler scaling factor and delay for severe doubly-spread time-variant channel. 
  
 \subsection{Path-specific  PTRM with Path-specific Channel Tracking  (PS-PTRM) }
If we  reverse the $\tau$ in \eqref{est_channel}, the output of PTRM with our channel tracking results as in \eqref{est_channel}  is 
\begin{equation}
\begin{aligned} 
{z}(t) &=  y(t)* {\hat h}( t;- \tau) \quad\quad \quad\quad  (k-1)T<t<kT,\quad k=1, 2, \cdots, \infty
\\
%
&= \sum\limits_{p = 1}^{\min\{{N_{\mathrm{pa}}}_k,\hat N_{\mathrm{pa}_k}\}} {{A_k^p}{{\hat A}_k^p}}  x((1 + {a_k^p})(1 - {\hat a_k^p})t + (1 + {a_k^p}){\hat \tau_k^p} - {\tau _k^p}) \\
&+ \sum\limits_{i = 1}^{{N_{\mathrm{pa}}}_k} {\sum\limits_{j = 1,j \ne i}^{\hat N_{\mathrm{pa}_k}} {{A_k^i}{{\hat A}_k^j} x((1 + {a_k^i})(1 - {{\hat a}_k^j})t + (1 + {a_k^i}){{\hat \tau }_k^j} - {\tau _k^i})} }\\
&\approx \sum\limits_{p = 1}^{\min\{{N_{\mathrm{pa}_k},\hat N_{\mathrm{pa}_k}\}}} \breve{A}_k^p  x([1 + \breve{a}_k^p]t + \breve{\tau}_k^p),   \end{aligned} 
\end{equation}
where 
\begin{subequations}
\begin{align}
\breve{A}_{k}^p &= {A_k^p}{\hat A_k^p},\\
\breve{a}_{k}^p&= {a_k^p} - {\hat a_k^p} - {a_k^p}{\hat a_k^p},\\
\breve{\tau}_{k}^p &= (1 + {a_k^p}){\hat \tau _k^p} - {\tau _k^p}.  
\end{align}
\end{subequations}

To analyze the performance, we consider that  the best case that $\hat N_{\mathrm{pa}_k}=N_{\mathrm{pa}_k}$, $\hat A_k^p=A_k^p$, $\hat a_k^p=a_k^p$ and $\hat \tau_k^p=\tau_k^p$,
then ${\hat h}( t; - \tau )={h}( t; - \tau )$.  The equivalent $Q$-function with the path-specific channel tracking results becomes $Q$-function in theory \eqref{qfunc}:
\begin{equation}\label{qfuncDC}
\begin{aligned} 
\hat{q}(t, \tau) &= {q}(t, \tau)   \quad\quad \quad\quad  (k-1)T<t<kT,\quad k=1, 2, \cdots, \infty \\
&= \sum\limits_{p = 1}^{{N_{\mathrm{pa}_k}}} {{A_k^p}^2} \delta (\tau  + {a_k^p}{\tau _k^p} - {a_k^p}^2t)\\
& +\sum\limits_{i = 1}^{{N_{\mathrm{pa}_k}}} {\sum\limits_{j = 1,j \ne i}^{{N_{\mathrm{pa}_k}}} {{A_k^i}{A_k^j}\delta (\tau  + [{a_k^i} - {a_k^j} - {a_k^i}{a_k^j}]t + {\tau _k^j} + {a_k^i}{\tau _k^j} - {\tau _k^i})} } .
\end{aligned} 
\end{equation}
Note that, this is different from conventional PTRM since conventional PTRM simply reverses the CIR in time domain,  not $\tau$ in  \eqref{est_channel}.
From \eqref{qfuncDC}, we can see that this PS-PTRM  can work well if  the Doppler dispersion is very small as $a_k^p\approx 0$. Then, with our channel tracking results, a more accurate $\hat \tau_k^p$ can help make \eqref{qfuncDC}  more focus to a delta function than conventional  PTRM for multipath time-variant channel. This can also work well if all paths have similar Doppler dispersion and we compensate the received data with one general  Doppler shift before the PTRM. However, for the severe doubly-spread channel \eqref{Dtchan},  Doppler scale $a_k^p$ can be too big and too different from each path,  we can compensate the Doppler and delay for each path with the following PTRM. 

\subsection{PTRM with Path-specific Compensation on Doppler  and delay  (PSC-PTRM) }

From the analysis in \eqref{qfuncDC},  we know if we can make the $Q$-function more focusing to  a delta function, it can perform better under severe doubly-spread channel.


Here we propose a PTRM with path-specific  compensation on Doppler shift and delay as follow. With 
$\{ {\hat A_k^p},{\hat \tau _k^p},{\hat a_k^p}\} _{p = 1}^{{\hat N_{\mathrm{pa}_k}}}$ for each time $k$  obtained by our channel tracking, the  channel with path-specific  compensation used for mirror  is as follow 
\begin{equation}
\hat h^\prime (t; \tau) = \sum\limits_{p = 1}^{{\hat N_{\mathrm{pa}_k}}} {{{\hat A}_k^p}\delta (\tau  - \hat   {\tau _k^p}^\prime + \hat   {a _k^p}^\prime t)}, \quad\quad \quad\quad  (k-1)T<t<kT,\quad k=1, 2, \cdots, \infty,
\end{equation}
where $  \hat   {\tau _k^p}^\prime= \frac{{{\hat \tau_k^p}}}{{1 + {\hat a_k^p}}}$, and $\hat   {a _k^p}^\prime = \frac{{{\hat a_k^p}}}{{1 + {\hat a_k^p}}}$. 


Then the output of PTRM with path-specific compensation on Doppler shift and delay is
\begin{equation}
\begin{aligned}
{ z'}(t) &= y(t)*\hat h' ( t;- \tau)  \quad\quad \quad\quad  (k-1)T<t<kT,\quad k=1, 2, \cdots, \infty\\
&= \sum\limits_{p = 1}^{\min\{N_{\mathrm{pa}_k}, \hat  N_{\mathrm{pa}_k}\}} {{A_k^p}{{\hat A}_k^p}} x(\frac{{1 + {a_k^p}}}{{1 + {\hat a_k^p}}}t + \frac{{1 + {a_k^p}}}{{1 + {\hat a_k^p}}}{\hat \tau _k^p} - {\tau _k^p}) \\
&+ \sum\limits_{i = 1}^{N_{\mathrm{pa}_k}} {\sum\limits_{j = 1,j \ne i}^{\hat N_{\mathrm{pa}_k}} {{A_k^i}{\hat A_k^j} x(\frac{{1 + {a_k^i}}}{{1 + {\hat a_k^j}}}(t + {\hat \tau_k^j}) - {\tau_k^i}) }}  \\
&\approx \sum\limits_{p = 1}^{\min\{N_{\mathrm{pa}_k}, \hat  N_{\mathrm{pa}_k}\}} {\check{A}_k^{p'}} x([1 + {\check{a}_k^{p'}}]t + {\check{\tau}_k^{p'}}),
\end{aligned} 
\end{equation}
where
\begin{subequations}
\begin{align}\label{1_a}
{\check  A_k^{p'}} &= {A_k^p}{\hat A_k^p},  \\
{\check a_k^{p'}} &= \frac{{{a_k^p} - {\hat a}_k^p}}{1 + \hat a_k^p},\\
{\check \tau_k^{p'}} &= \frac{{1 + {a_k^p}}}{{1 + {\hat a_k^p}}}{\hat \tau _k^p} - {\tau _k^p}.
\end{align}
\end{subequations}
Obviously, in the best case for estimation, if $\hat N_{\mathrm{pa}_k}=N_{\mathrm{pa}_k}$, $\hat A_k^p=A_k^p$, $\hat a_k^p=a_k^p$ and $\hat \tau_k^p=\tau_k^p$, the energy can be focus on the mainlobe.
In this PTRM with path-specific compensation on Doppler scaling factor and delay, the $Q$-function becomes
\begin{equation}\label{qfuncDCv}
\begin{aligned} 
{ q'}(t,\tau) 
&= \sum\limits_{p = 1}^{{N_{\mathrm{pa}_k}}} {{A_k^p}^2} \delta (\tau)\\
& + \sum\limits_{i = 1}^{{N_{\mathrm{pa}_k}}} {\sum\limits_{j = 1,j \ne i}^{{N_{\mathrm{pa}_k}}} {{A_k^i}{A_k^j}\delta (\tau  + \frac{{{a_k^i} - {a_k^j}}}{{1 + {a_k^j}}}t + \frac{{1 + {a_k^i}}}{{1 + {a_k^j}}}{\tau _k^j} - {\tau _k^i})} }\\
&\approx \sum\limits_{p = 1}^{{N_{\mathrm{pa}_k}}} {{A_k^p}^2} \delta (\tau),\\
& \quad  (k-1)T<t<kT,\quad k=1, 2, \cdots, \infty.
\end{aligned} 
\end{equation}

Comparing with \eqref{qfuncDC}, the mainlobe of \eqref{qfuncDCv} is   a delta function. This makes the energy  more focusing, thus this PSC-PTRM can work better in our fast time-variant, and severe doubly-spread channel.


\section{Simulation}

In our simulations, suppose that the receiver and transmitter are located in the same water depth of shallow water in Fig. \ref{fig:channnel1}. The depth of receiver ${h_{10}} = 50m$, the distance between receiver and the bottom ${h_{20}} = 100m$, the initial distance between receiver and transmitter $d_{\mathrm{SR}} = 500m$, and the speed of transmitter $v = -5m/s$ (When the receiver and transmitter move to each other, $v>0$). The receiver is still.  And the speed of sound  $c = 1500m/s$.

To simulate the doubly-spread channel, besides the relative moving of the receiver, we generate five main paths including direct path, and four other paths reflecting from surface and bottom as shown in Fig.\ref{fig:channnel1}. The amplitudes for paths are assumed to be the geometric spreading, which is a hybrid of spherical and cylindrical spreading, with the power loss to be proportional to ${d^\beta }$ where $d$ is  the distance of the path and $\beta$ is between 1, for cylindrical spreading, and 2, for spherical spreading. Provided that the sound propagation in real channel can hardly be classified into either of the two spreading models, a practical value of the spreading exponent can be taken as $\beta=1.5$.  
\begin{figure}[htbp]
	\centering
	\includegraphics[angle=0,width=0.9\textwidth]{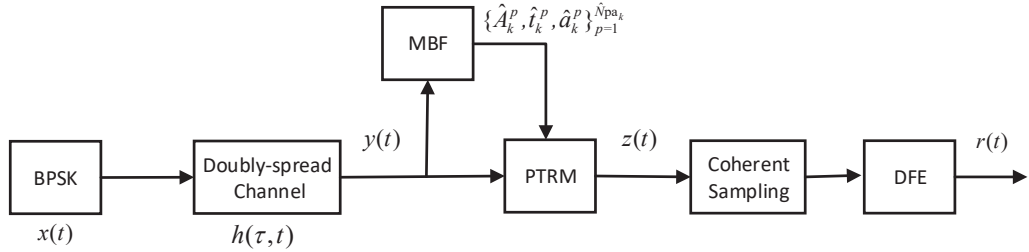}
	\caption{The block diagram of our underwater acoustic  communications from the transmitter to the receiver.}
   \label{system}
\end{figure}

Our communication system block diagram is shown in Fig.\ref{system}. 
We use BPSK single carrier system in the simulation. Denote $ x(t)$ as the transmitted passband signal. The received passband signal through the doubly-spread underwater acoustic channel  is $y(t)$.  After we receive $ y(t)$, our path-specific MB channel tracking is applied to track the channel. Then with our proposed path-specific PTRMs  we obtain $ z(t)$.  After  coherent demodulation and DFE, we arrive to $r(t)$.

\begin{figure}[htbp]
	\centering
	\includegraphics[angle=0,width=0.7\textwidth]{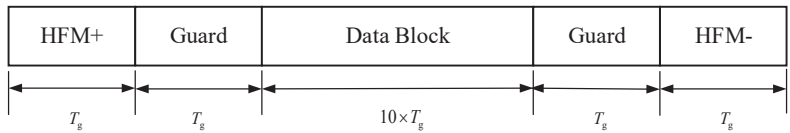}
	\caption{The sending data block $ x(t)$.}
   \label{signal}
\end{figure}

The sending signal block $ x(t)$ for communications consists of preamble signals, guard intervals and data as in Fig.\ref{signal}.   In this  simulation, hyperbolic frequency modulated (HFM)  signals are chosen as the probe signals. HFM is widely used as a preamble signal in underwater acoustic communications \cite{Shengli2014}.  HFM+ is the up-sweep HFM, while HFM- is the down-sweep HFM. With them, we can obtain multipath measurement set $\mathbf{Z}_k(k=1,2,...)$ defined in \eqref{measset} for channel tracking following the methods in \cite{Sharif} \cite{MHFM}. Both  HFM signal and guard intervals have the same length ${T_\mathrm{g}}=100ms$. 
For the transmitted signal the centering frequency ${f_\mathrm{c}}= 5\mathrm{kHz}$, sampling frequency ${f_\mathrm{s}} = 50\mathrm{kHz}$, and with  $\mathrm{SNR}=5 \mathrm{dB}$.

\begin{figure}[htbp]
\label{results}
\centering
\subfigure[Channel tracking results of delay: $\cdot$  true delay of paths, {\color{red}$\circ$}  the measurements of the delay of paths, {\color{blue}$+$} the tracked delay of paths ]
{\label{fig:ep:a}\includegraphics[width=0.45\textwidth]{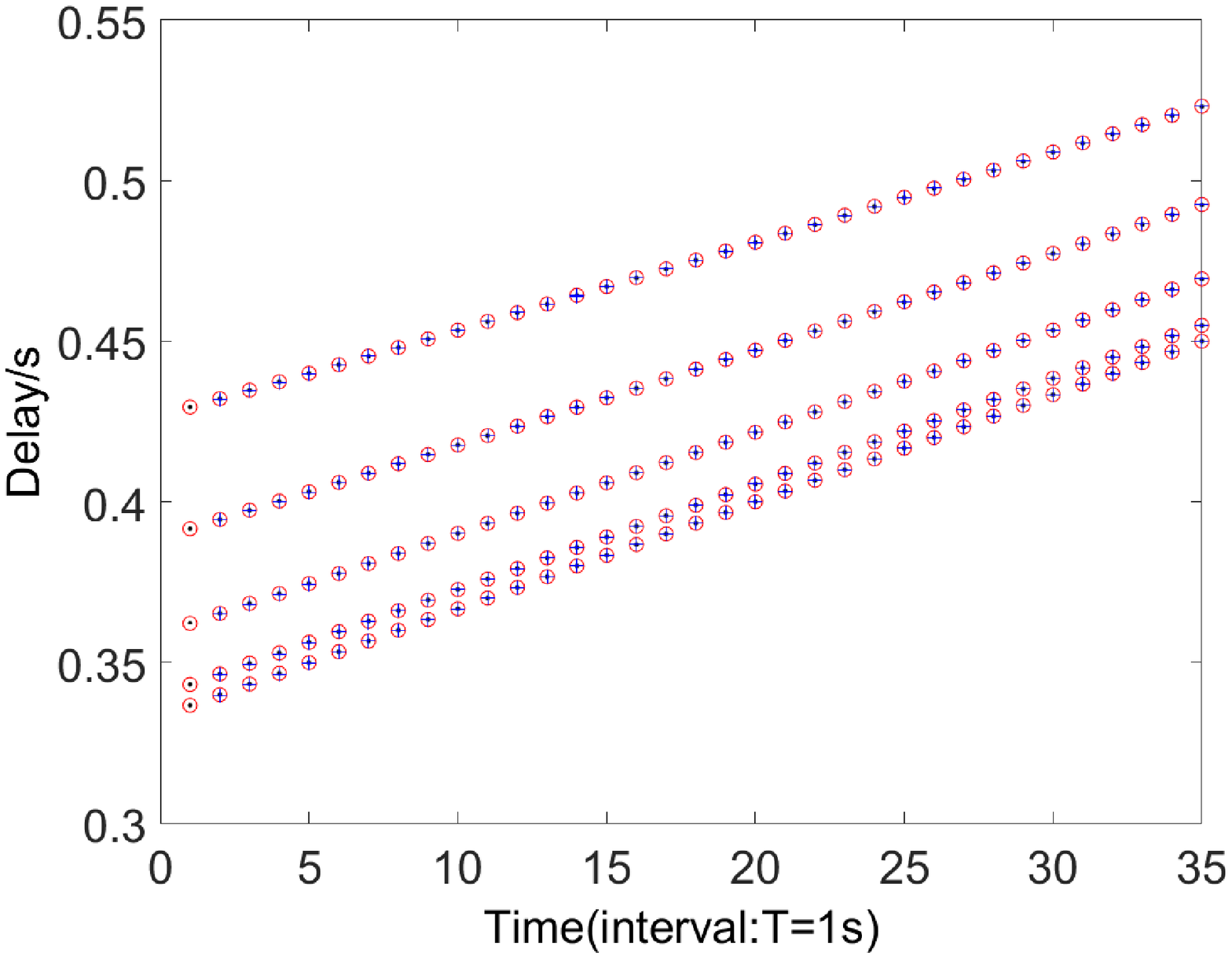}}
\hspace{-0in}
\subfigure[Channel tracking results of Doppler scaling factor: $\cdot$  true Doppler scaling factor, {\color{red}$\circ$}  the measurements of the Doppler scaling factor, {\color{blue}$+$} the tracked Doppler scaling factor]
{\label{fig:ep:b}\includegraphics[width=0.45\textwidth]{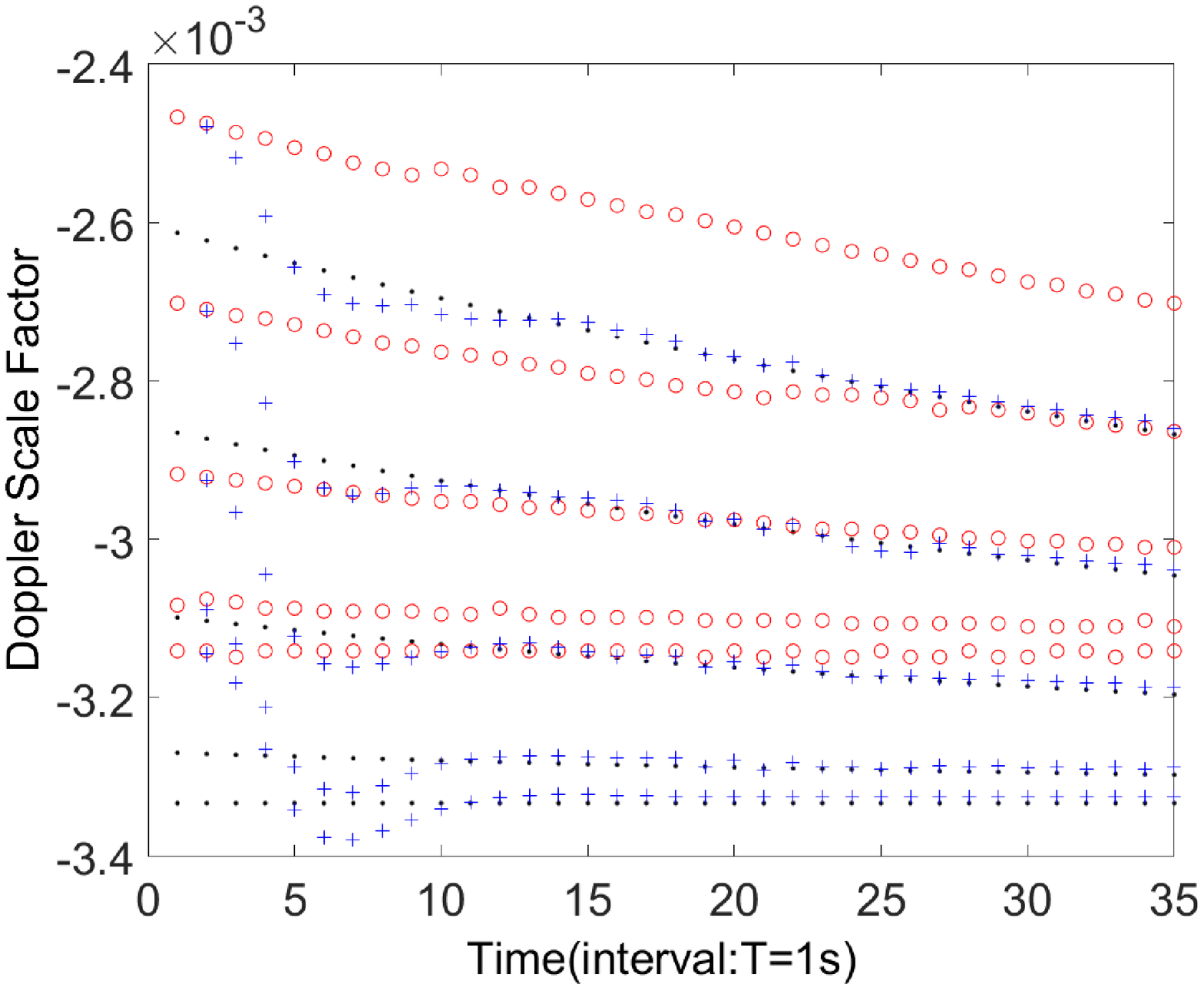}}
\subfigure[ OSPA metric of the measurements and the path-specific channel tracking. ]
{\label{fig:ep:c}\includegraphics[width=0.45\textwidth]{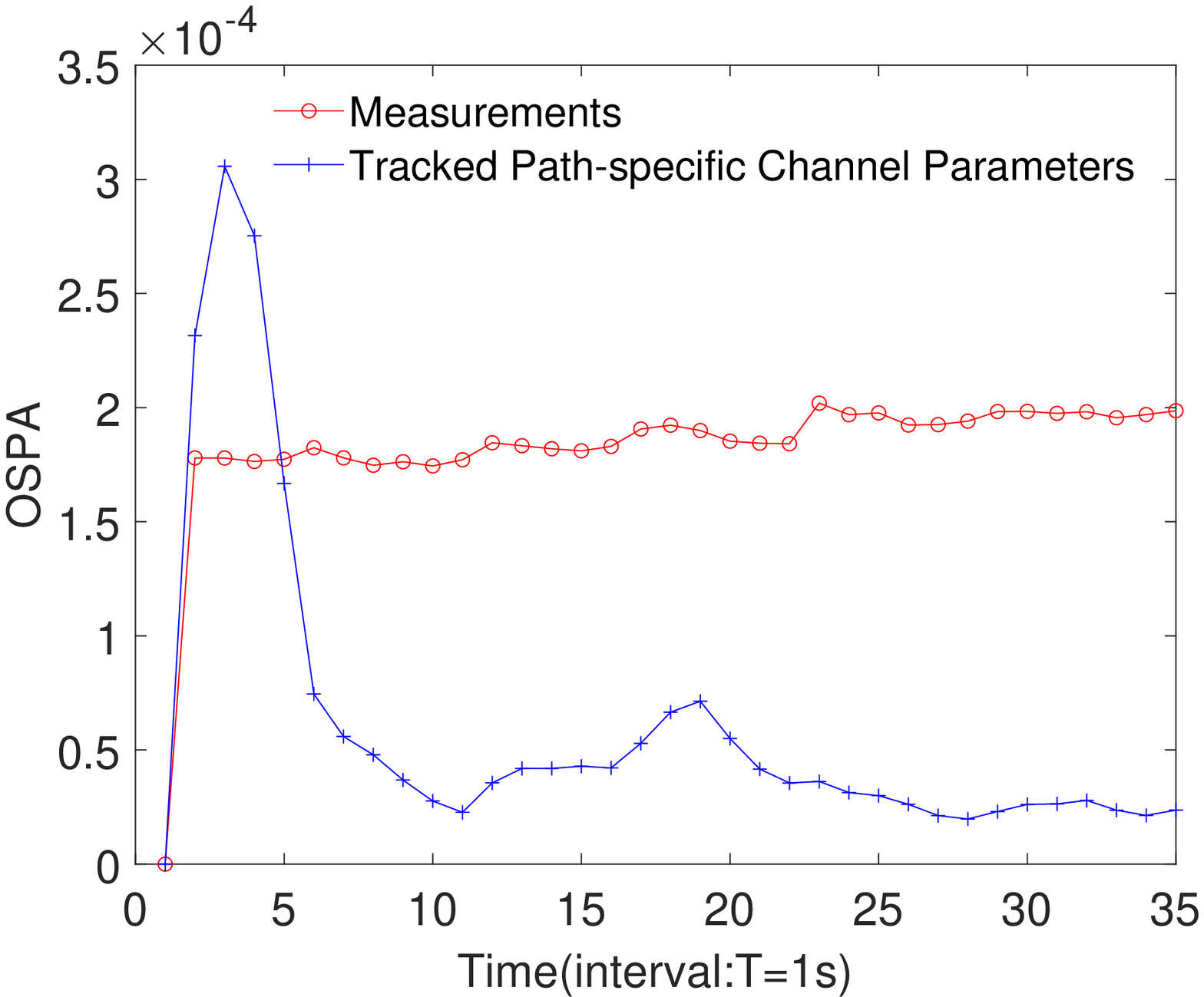}}
\caption{Path-specific channel tracking results by simulated data. The time state interval $T=1\rm s$ in our simulation to track the channel. The measurements are obtained by HFM+ and HFM-  \cite{Sharif} \cite{MHFM}.}
\end{figure}

The tracking results of delay and Doppler scaling factor for paths in one block time are demonstrated in Fig.\ref{fig:ep:a} and Fig.\ref{fig:ep:b} respectively. The measurements for trackers are extracted from a rough channel estimation by preamble signals HFM+ and HFM-, and shown  in Fig.5. The time interval of each tracking state is set as $T = 1 \rm s$.
From Fig.\ref{fig:ep:a} we show that the delay scale is from around $0.35 \rm s$ to around $0.53\rm s$. And the scale of Doppler scaling factor is from around $-3.4 \times {10^{{\mathrm{ - 3}}}}$ to around $-2.6 \times {10^{{\mathrm{ - 3}}}}$. These demonstrate the time-variant and doubly-spread  property of the channel. With methods \cite{Sharif} \cite{MHFM}, we can have the rough measurements of the delay and Doppler at each time. 
For both delay and Doppler scaling factor, the tracking results are closer to the real parameters we set in the simulations than the measurements with time. This means that the channel tracking improve the acquiring  of channel information especially for Doppler scaling factor in this simulation.


We also provide the optimal sub pattern assignment (OSPA) metric to show the tracking performance in Fig.\ref{fig:ep:c}. 
OSPA metric is an important metric for tracking. It is defined as \cite{Schuhmacher}
\begin{align}
\label{eqn_OSPA}
&\bar{d}^{(c)}_p(\mathbf{X},\mathbf{Y})=\nonumber\\
&\left(\frac {1}{n}\left(\min_{\pi\in\Pi_n} \sum_{i=1}^m d^{(c)} (\mathbf{x}_i,\mathbf{y}_{\pi(i)})^p + c^p(n-m)\right)\right)^{1/2},
\end{align}
if $m\le n$, and $\bar{d}^{(c)}_2(\mathbf{X},\mathbf{Y}):=\bar{d}^{(c)}_2(\mathbf{Y},\mathbf{X})$ if $m>n$; 
is the Euclidean distance between estimates $\mathbf{x}$ and real parameters $\mathbf{y}$ cut off at $c$, and $\Pi_n$ is the set of permutations on $\{1,2,...n\}$ to show the correspondence between estimated paths and real paths.


OSPA results of both measurements and results  of proposed channel  tracking are shown in Fig.\ref{fig:ep:c}. For the first several time states, the tracker has a higher OSPA than measurements, but after several time states, it shows a good convergence comparing to the measurements. 
 
\begin{figure}[htbp]
\centering
\subfigure[MSE of delay;]
{\label{MSE1}\includegraphics[width=0.45\textwidth]{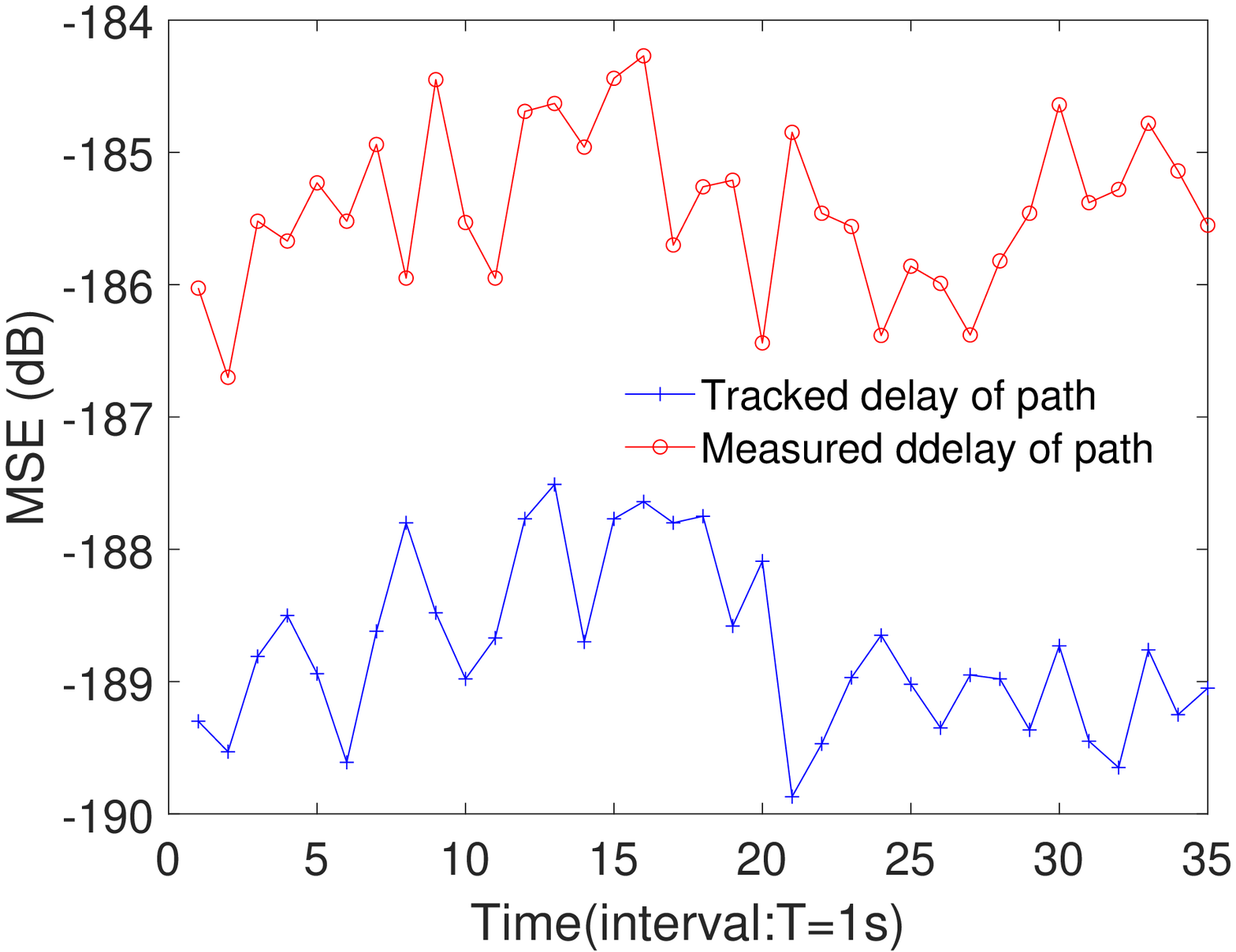}}
\hspace{-0in}
\subfigure[MSE of Doppler scaling factor;]
{\label{MSE2}\includegraphics[width=0.45\textwidth]{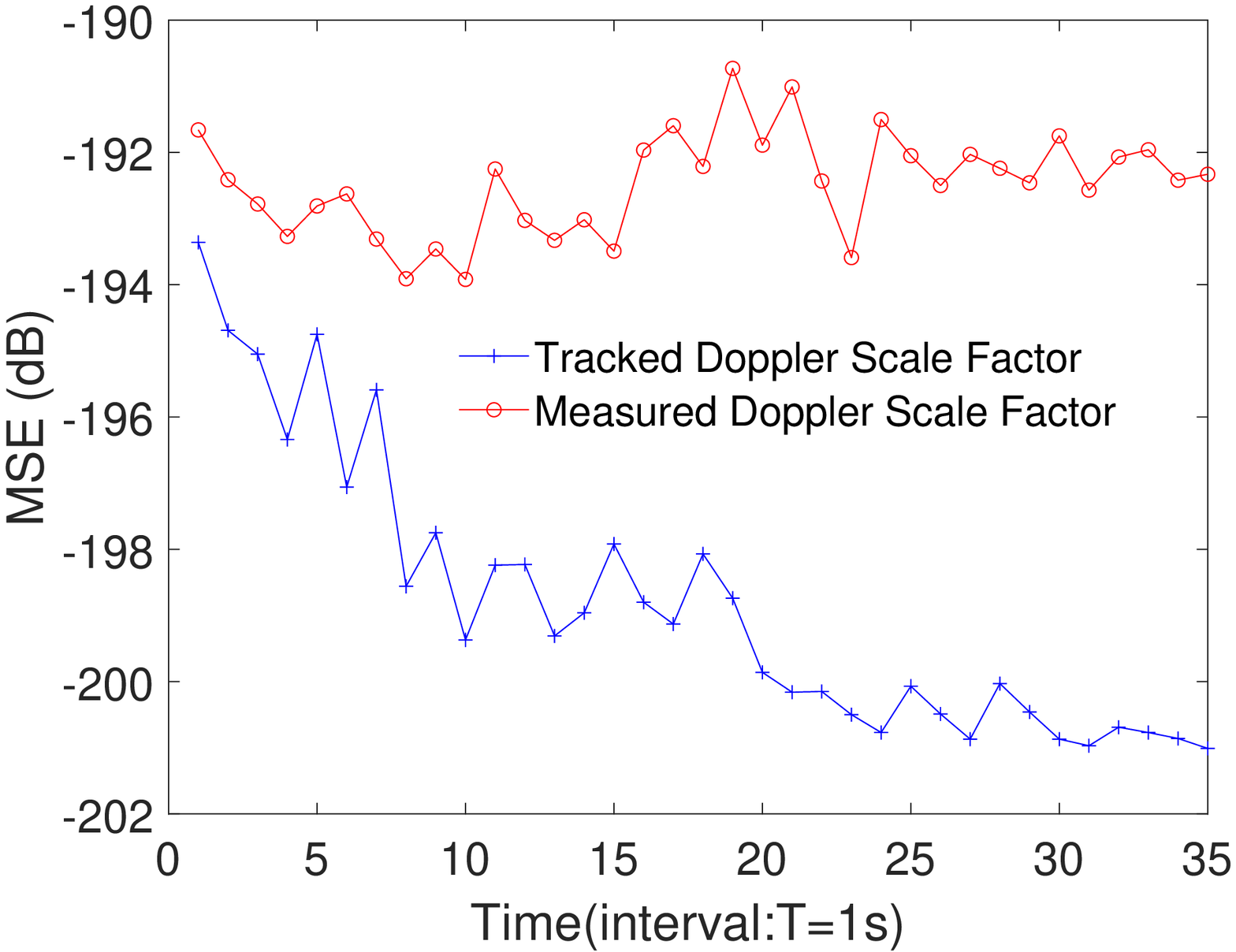}}
\caption{MSE for  path-specific channel tracking results on delay and Doppler scaling factor in simulation.}\emph{}
\end{figure}

Fig. 6  provides the MSE of measurements and  our  proposed multipath channel tracking after 1000 times of Monte Carlo simulations.  
Generally, after our proposed channel tracking, the MSE becomes  lower than measurements for both delay and Doppler.
With longer tracking time, the error of tracking results become less especially for Doppler.

\begin{figure}[htbp]
	\centering
	\includegraphics[angle=0,width=0.45\textwidth]{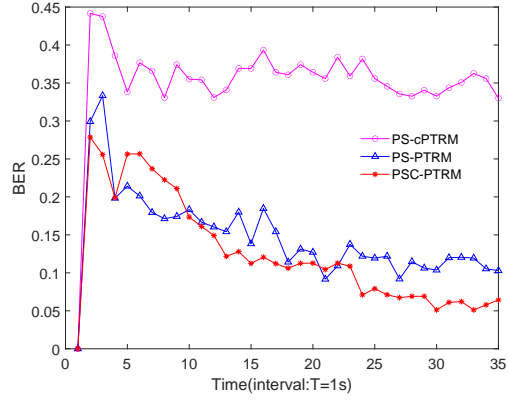}
	\caption{BER of different  PTRMs with path-specific channel tracking. PS-cPTRM is the conventional PTRM with CIR by the path-specific channel parameters; PS-PTRM and PSC-PTRM are the two proposed PTRM with path-specific channel tracking results. }
   \label{BER}
\end{figure}

We also provide the BER performance in our simulations in Fig.\ref{BER}. BER is  crucial for the communications,  it can demonstrate how much  our proposed methods can improve the overall system. We apply our proposed channel tracking as in Fig.\ref{fig:ep:a} and Fig.\ref{fig:ep:b} to  our proposed PTRMs and  the conventional PTRM \cite{TCSPTRM}. 
Fig.\ref{BER} shows that the path-specific channel trackers can be applied in underwater acoustic communications.
Both proposed PTRMs  have much lower BER than Conventional PTRM with CIR obtained by the path-specific channel parameters. And BER converges with time by our proposed methods. This means our methods can efficiently  improve the communication performance. 
PSC-PTRM  has a little lower BER than PS-PTRM in Fig.\ref{BER} with longer time steps since with path specific compensation for double-spread channel, we can improve the performance further. PSC-PTRM does not improve too much comparing to PS-PTRM  in this simulation because that it works better if the frequency dispersion is more intensive  which means Doppler scaling factors are bigger and different. In our simulations, with the time grows, the Doppler scaling factors grow bigger , therefore the BER converge with time  in Fig.\ref{BER}.

\section{Experiment}

\begin{figure}[htbp]
	\centering
	\includegraphics[angle=0,width=0.6\textwidth]{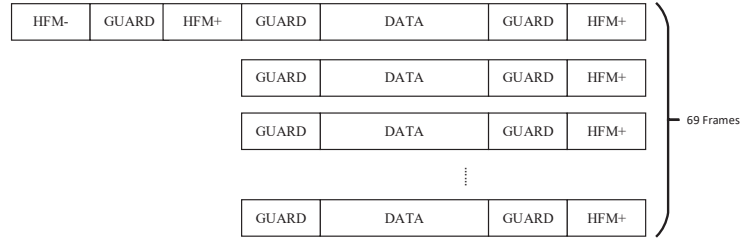}
	\caption{The sending data block in Qiandao Lake experiment 2016.}
	\label{data}
\end{figure}

\begin{figure}[htbp]
	\centering
	\includegraphics[angle=0,width=0.45\textwidth]{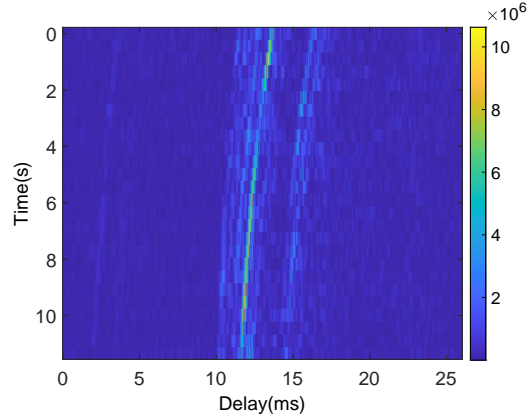}
	\caption{Measured CIRs over 26 consecutive frames in Qiandao Lake experiment 2016.}
	\label{CIR}
\end{figure}

The Qiandao Lake experiment was conducted in Qiandao Lake, Zhejiang on May 3, 2016. There are 4 receivers in a vertical array. The sound speed is nearly a constant for these four receivers. The source moves away from receiver, and the relative speed  is $v = -5.14\mathrm{ m/s}$. Some experimental parameters are set as follow: center frequency
$f_{\mathrm c }= 12{\mathrm{kHz}}$, sampling frequency $f_{\mathrm s} = 96{\mathrm{ kHz}}$, symbol rate of QPSK $R_{\mathrm s} = 6{\mathrm{kBaud}}$, time duration of data frame $T_{\mathrm f} = 389.33{\mathrm{ms}}$, time duration of HFM and guard interval $T_{\mathrm g} = 21.33{\mathrm{ms}}$. There are 69 data frames shown in Fig.\ref{data}.  

Before and after every data frame, there is an HFM+ respectively, which is used for frame synchronization. We also use them for Doppler estimation. There is only one HFM- signal at the very beginning of the data. 
 The measured CIR by HFM+ signals over 26 consecutive frames in Qiandao Lake experiment are shown in Fig.\ref{CIR}. We can extract  relative channel parameters from them as the measurements, and feed them to trackers. 
 
 Some other parameters of MB trackers  are set as follow in this experiment:  the probability of survival $P_{\mathrm s} = 0.999$, observation interval $T = 453.3\mathrm{ms}$, processing noise  covariance 
 $\mathbf{Q}_k=\mathrm{diag} \left[10^{- 4}, 10^{- 6}\right]$, 
 measurement error covariance $\mathbf{R}_k =\mathrm{diag} \left[ 
10^{-5}, 10^{-6}\right]$, pruning threshold $\tau_{\mathrm p} = 10^{-4}$, confirming threshold $\tau_{\mathrm c} = 0.75$, existence threshold $\tau_{\mathrm e} =0.25$.

\begin{figure}[htbp]{\label{QL}
\centering
\subfigure[Channel tracking results of delay;]
{\label{delay}\includegraphics[width=0.45\textwidth]{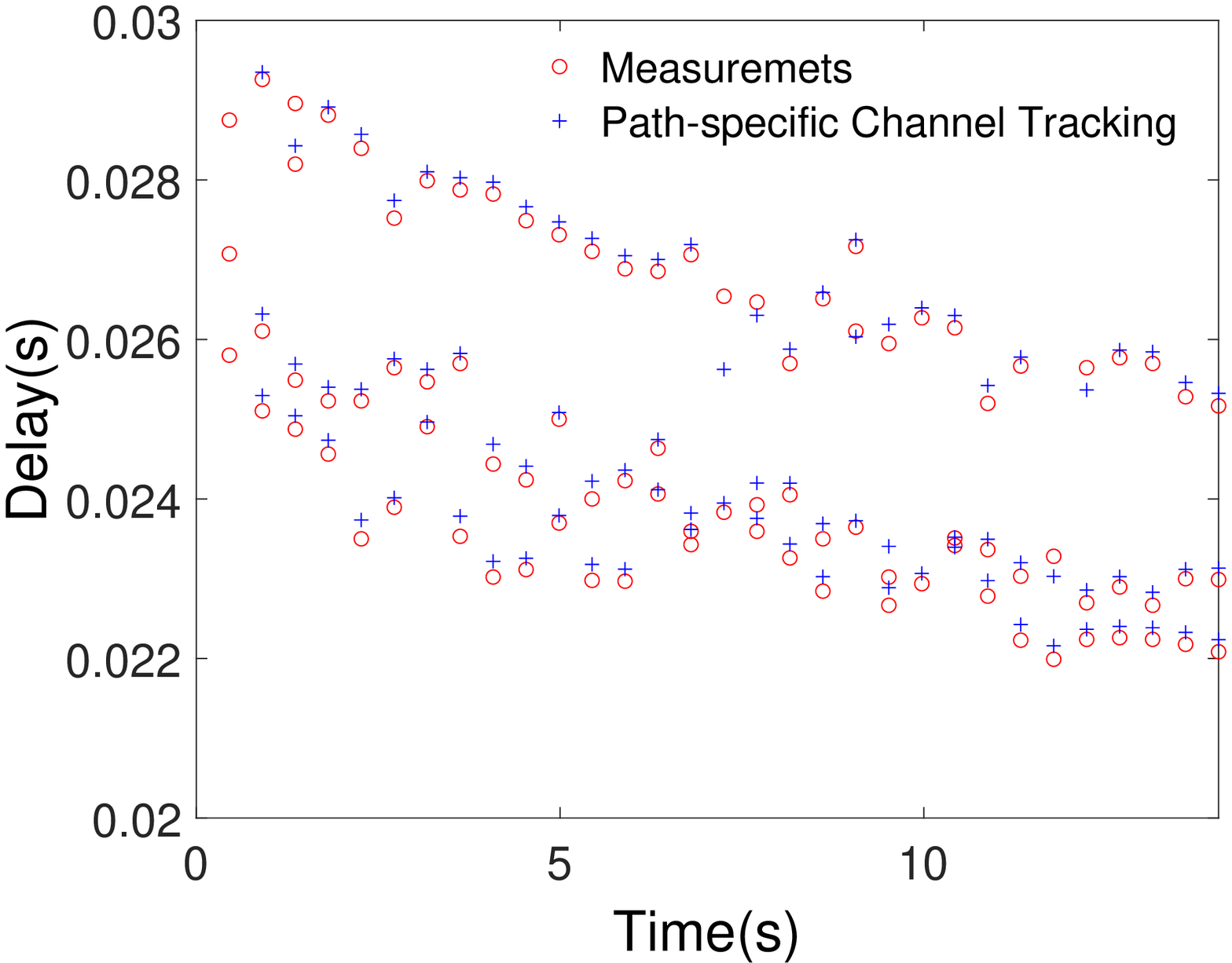}}
\hspace{-0in}
\subfigure[Channel tracking results of Doppler scaling factor;]
{\label{doppler}\includegraphics[width=0.45\textwidth]{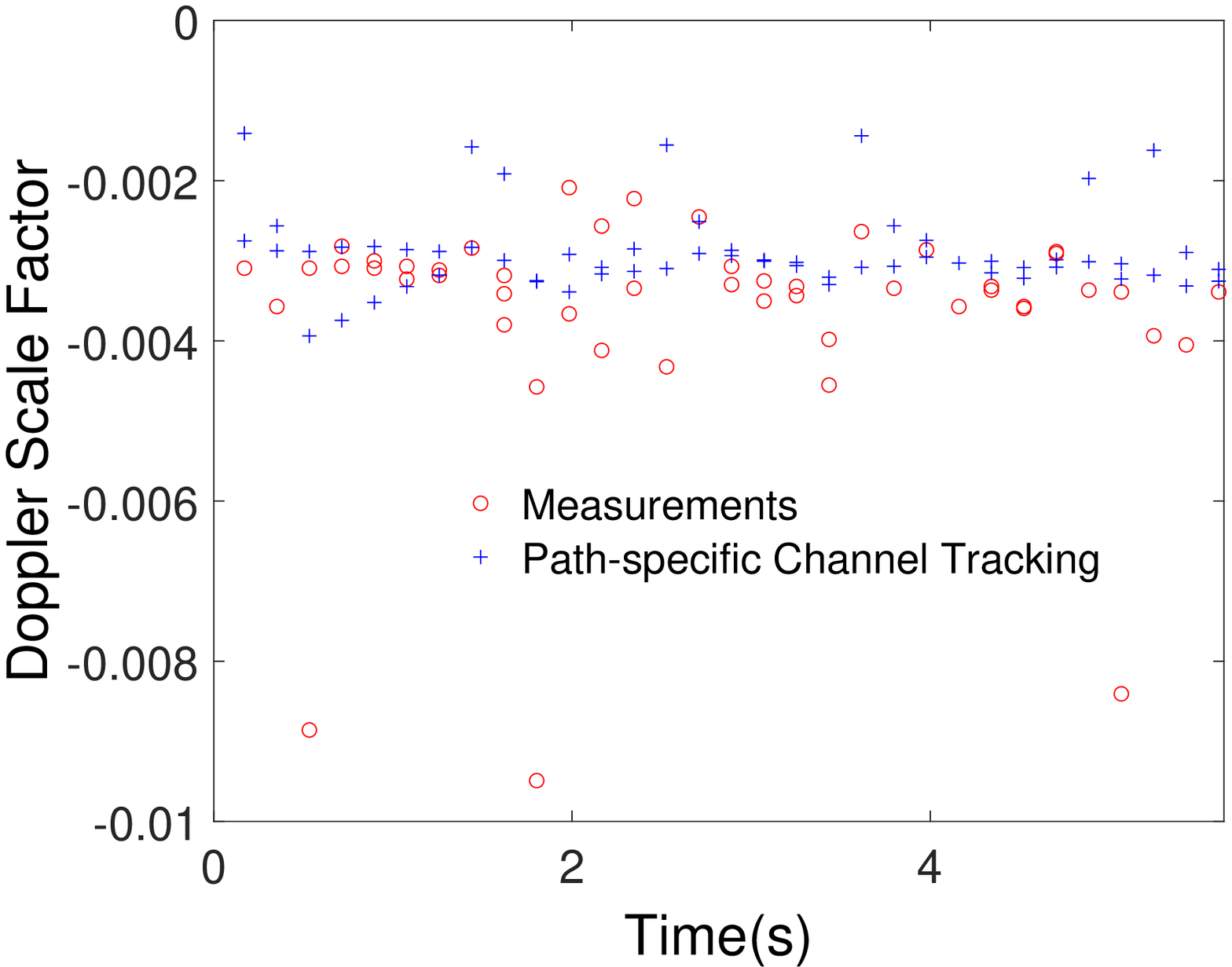}}
\caption{Path-specific channel tracking results over 26 consecutive frames in Qiandao Lake 2016 experiment.}
}
\end{figure}
\begin{figure}[htbp]
	\centering
	\includegraphics[angle=0,width=0.45\textwidth]{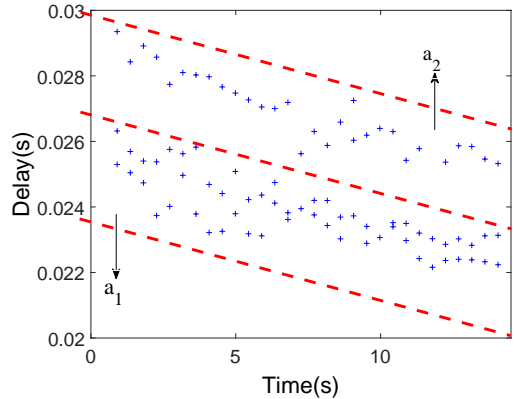}
	\caption{The assignment of the tracked  Doppler scaling factor to the tracked delay.}
	\label{assign}
\end{figure}

The path-specific channel tracking results are shown in Fig.10. 
In Fig.\ref{delay} the MB trackers track three paths during tracking  delay. 
Two paths have similar and smaller delay. One single path is with longer delay.
In Fig.\ref{doppler},  only two Doppler scaling factors in a frame are captured.  This  is because  that  Doppler scaling factors are  smaller than delay in scale and  some Doppler scaling factors for different paths are too close.
Fig.\ref{assign} shows how to assign two specific Doppler scaling factors to the three  delays. According to our sound propagation model in Fig. \ref{fig:channnel1} and Fig.\ref{fig:channnel2}, paths with smaller delay should have larger Doppler, therefore,
we assign smaller Doppler $a_2$ to  the single path with longer delay,  and the bigger $a_1$ to the two paths with  similar smaller delay  in Fig.\ref{doppler}.

Fig.12 demonstrates the BERs of the Qiandao Lake experiment with different methods.
The  conventional PTRM with channel estimation by least square (LS) doesn't work well even if with DFE.
The  conventional PTRM with the measurements in this work without tracking procedure is slightly better than LS, since more parameters are prepared.
Clearly, MB-based path-specific channel tracking improves the accuracy of estimated CIR  hence improves the communication performance. 
PSC-PTRM has the lowest BER in this experiment. This is because the conventional PRTM is only 
for time-invariant channel. But for real underwater acoustic communications, the channel becomes doubly spread channel, Doppler is inevitable due to the physical property of water.  After we obtain Doppler scaling factors for different paths by our proposed channel tracking and carry out path-specific Doppler compensation, the BER is further reduced.  The convergence of the MB based path-specific channel tracking prove the stability of the tracking.

\section{Conclusion}
In this work, we propose a path-specific underwater acoustic channel tracking, and path-specific PTRMs with the proposed channel tracking. With a sound propagation model, we provide the state transition model for 
time-variation of channel. And with the MB-tracker's help we can track the delay and Doppler for each path and the multipath number.
Then, we extend the conventional PTRM algorithm to time-variant and doubly-spread channel  with  path-specific parameters we obtain by the proposed channel tracking. 
The simulation and Qiandao Lake experiment are based on single carrier communication system. 
Both show the efficiency of the path-specific channel tracking and the proposed PTRMs. The experiment shows that even with some practical 
model-mismatch in real water, the proposed method is still robust.
In fact, this path-specific underwater acoustic channel tracking is not restricted to any modulation, and has much potential to improve the underwater acoustic communications and signal processing. Our further work will show the detection/location results with the proposed path-specific  channel tracking.

\begin{figure}[htbp]
	\centering
	\includegraphics[angle=0,width=0.7\textwidth]{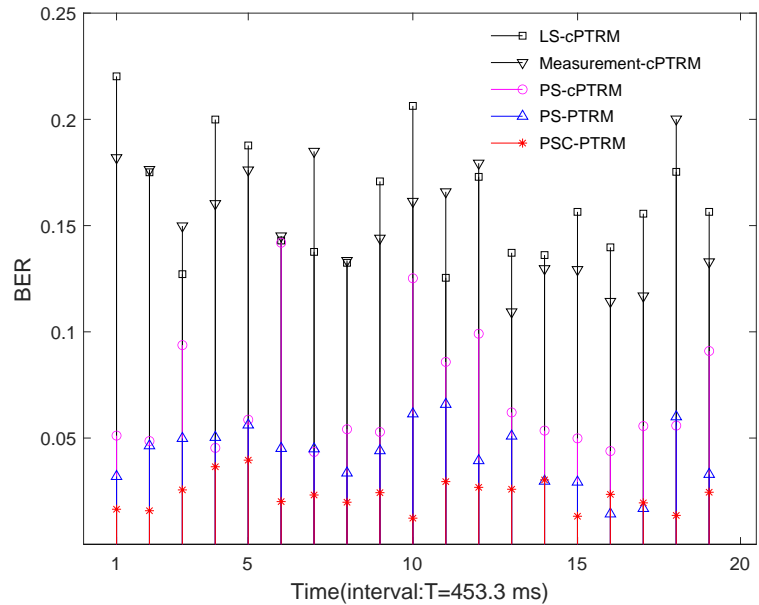}
	\caption{BER results of different methods. 
	 LS-cPTRM denotes the BER is obtained with channel estimation by LS method, then  conventional PTRM.
The measurement-cPTRM represents our measurements  and conventional PTRM. }
	\label{QL1}
\end{figure}

\section*{Acknowledgment}
Thank Prof. Lijun Xu for providing the 2016 Qiandao Lake experiment data.

\end{document}